\documentclass{PoS}

\usepackage{xspace,url,amsmath,amssymb,booktabs}

\newcommand*\diff{\mathop{}\!\mathrm{d}} % d for integrals
\newcommand{\ptwo}{\textit{P2}\xspace}
\newcommand{\pone}{\textit{P1}\xspace}

\newcommand{\eqgn}{E_{\textit{QG}_n}\xspace}
\newcommand{\eqgone}{E_{\textit{QG}_1}\xspace}
\newcommand{\eqgtwo}{E_{\textit{QG}_2}\xspace}
\newcommand{\epr}{E^\prime} % for average
\newcommand{\phipr}{\phi^\prime} % for average
\newcommand{\bonu}{{\boldsymbol \nu}} % for average
\newcommand{\phip}{\phi_\textit{\!P2}} % for average
\newcommand{\sigmap}{\sigma_\textit{\!P2}} % for average
\newcommand{\hphip}{\widehat{\phi}_\textit{\!P2}} % for average
\newcommand{\hsigmap}{\widehat{\sigma}_\textit{\!P2}} % for average

\title{Constraining Lorentz invariance violations using the Crab pulsar TeV emission}

\ShortTitle{LIV from Crab pulsar TeV emission}

\author{\speaker{Markus Gaug}\\
 Unitat de F\'isica de les Radiacions, Departament de F\'isica, and CERES-IEEC, Universitat Aut\`onoma de Barcelona, E-08193 Bellaterra, Spain\\
        E-mail: \email{markus.gaug@uab.cat}}

\author{Daniel Garrido\\
 Unitat de F\'isica de les Radiacions, Departament de F\'isica, and CERES-IEEC, Universitat Aut\`onoma de Barcelona, E-08193 Bellaterra, Spain\\
        E-mail: \email{daniel.garrido@uab.cat}}

\abstract{Fast variations of gamma-ray flux from Active Galactic Nuclei and Gamma-Ray Bursts can constrain Lorentz Invariance Violation (LIV) 
because of the delayed (or advanced) arrival of photons with higher energies: 
this approach has lead to the current world-best limits on the energy scale of Quantum Gravity. 
Here we report on constraints on LIV studying the gamma-ray emission up to TeV energies from the Galactic Crab pulsar, recently discovered by the MAGIC collaboration. 
A likelihood analysis of the pulsar events reconstructed for energies above 400 GeV finds no significant variation of energy-dependent arrival time, 
and 95\% CL limits are then obtained on the effective LIV energy scale after taking into account systematic uncertainties. 
Only a factor of about two less constraining than the current world-best limit on a quadratic LIV scenario, 
pulsars are now well established as a third and independent class of astrophysical objects suitable to constrain the characteristic energy scale of LIV.}

\FullConference{35th International Cosmic Ray Conference - ICRC217-\\
		10-20 July, 2017\\
		Bexco, Busan, Korea}

\begin{document}

\section{Introduction}

Quantum Gravity (QG) models \cite{rovelli04} combine the field equations of general relativity with quantum field theory. 
Many such scenarios include spontaneous violation of the Lorentz invariance (LIV)~\cite{Kostelecky1989,Burgess2002,lqgoptics,Douglas2001,Magueijo:2002,HamedArkani2004,Horava:2009}, 
which lead, among others, to energy dependent dispersion relations of the photon in vaccuum. While such effects are expected to become important at energies 
of the order of the Planck scale ($E_{\textit{Pl}} = \sqrt{\hbar	c^5 / G} \approx 1.22 \cdot 10^{19}$~GeV), they can manifest themselves already at much lower energies, through 
tiny deviations from the Lorentz invariant scenario, which accumulate once the photons travel very large distances~\cite{amelino2009}. 
Introducing an effective quantum gravity scale $\eqgn$, which may be of the order of the Planck energy or lower, 
the group velocity of photons of energy $E \ll \eqgn$ can be expressed as an expansion in powers of $E$  (see e.g.~\cite{amelino2009}), where: %  
\begin{equation}\label{eq.livphotonspeed}
	u_{\gamma}(E) = \frac{\partial E}{\partial p} \approx c \cdot \left[ 1 - \xi_n \frac{n+1}{2} \left( \frac{E}{\eqgn}\right)^n \right]  \quad.
\end{equation}
Here, $\xi_n = +1$ stands for a subluminal scenario, while $\xi_n = -1$ characterizes a superluminal
scenario, and $\xi_n =0$ for the case that the corresponding order is forbidden\footnote{% 
Eq.~\ref{eq.livphotonspeed} neglects terms breaking rotation invariance
which if there, would however imply some breaking of boost invariance as well~\cite{livtests2005}}.
We consider here terms with $n > 0$, which produce energy dependent velocities, typically considered in time-of-flight experiments, and there 
the linear case of $n=1$ and the quadratic case $n=2$.
Odd terms of $n$ violate \textit{CPT}~\cite{colladay1998}, that's why the $n=2$ case may dominate if \textit{CPT} is conserved. 

Energy-dependent arrival time variations have been studied so far using flares from Active Galactic Nuclei (AGNs)~\cite{liv-mkn501,hessliv2011},  %\cite{agn93}
and the very fast flux variations of Gamma-ray bursts (GRBs)~\cite{fermiliv13,amelino98}.  %\cite{grbdiscovery}
The latter have achieved  sensitivities to the linear case $\eqgone$ of well beyond the  Planck scale~\cite{fermiliv13,Goetz:2014} which has 
been effecitvely excluded. 
Constraints on $\eqgone$ have been obtained from the Crab pulsar starting from 1969 already~\cite{craborig1969} and constantly improved since 
then~\cite{kaaretcrab,nepomuk}.
Although the Crab pulsar is found many orders of magnitude closer to us than AGNs and GRBs, some of them observed at cosmological distances, 
its pulsations repeat and can be added over many periods to improve sensitivity to LIV. 

\section{MAGIC observation of TeV emission from the Crab Pulsar}

The Major Atmospheric Gamma-ray Imaging Cherenkov system (MAGIC) is located at the Roque de los
Muchachos observatory ($28.8^{\circ}$N, $17.8^{\circ}$W, 2200~m a.s.l.), in the Canary Island of La Palma, Spain.
The MAGIC system consisted of a single 17~m-dish telescope during its first 5 years of operation~\cite{magicperform} 
to which, in 2009, a second telescope was added in order to create a stereo system~\cite{perform12}.
A major upgrade was carried out between 2011 and 2012~\cite{perform13,perform16}. 
%In stereoscopic observation mode, the system reaches a maximum sensitivity of $\sim 0.6$\% of the Crab Nebula flux,
%for energies above $\sim 300$~GeV in 50 hours of observation~\cite{perform16}. % \cite{magic16}. % citation not needed, same as previous one

The Crab Nebula, together with its Pulsar, which cannot be spatially separated so far by Imaging Atmospheric Cherenkov Telescopes (IACTs), 
has been observed by MAGIC in every possible hardware configuration since its start. 
Being the brightest steady Very High Energy (VHE) gamma-ray source in the sky, it is 
regularly observed for calibration purposes and performance monitoring, leading to more than thousand hours of total observation time. 
These data have been down-selected to slightly more than 300~h of excellent quality, including single telescope (``mono'') and 
dual telescope (``stereo'') configurations, and requiring simultaneous precision time stamping 
which allows to attribute a precise pulsar phase value to each registered event, using 
ephemeres provided by the Jodrell Bank Observatory~\cite{jodrellbank}. 

With these data, MAGIC has detected emission from the Crab Pulsar up to 0.5~TeV for the main pulse \pone, and up to 1.5~TeV for the inter-pulse
\ptwo \cite{crabtev}, showing $544 \pm 92$ excess events for \ptwo above 400~GeV. The spectrum of both pulses is consistent with a power-law, 
however a significant difference was found between the reconstructed spectral indices of \pone and \ptwo, the latter being harder~\cite{crabtev}. 
This unique set of data is now used to test LIV. 

\section{Maximum likelihood method}

We construct a maximum likelihood method following the approach of~\cite{manelmanel}, further elaborated in~\cite{hessliv2011,fermiliv13}, and slightly 
adapted for the quadratic LIV effect. %Its main advantage is that it uses
Two new parameters are defined: $\lambda_1 \equiv 10^{19}~\mathrm{GeV}/\eqgone$ and
$\lambda_2 \equiv 10^{12}~\mathrm{GeV}/\eqgtwo$. 
The LIV effect under test (Eq.~\ref{eq.livphotonspeed}) produces then a mean phase delay of\,\footnote{%
The definition of $\lambda_2$ differs slightly from~\cite{hessliv2011,fermiliv13}, 
which is now directly proportional to $1/\eqgtwo$ (the quantity of interest), instead of $1/E^2_{\textit{QG}_2}$.}:
\begin{eqnarray}
\Delta\phi_n &=& c_n \cdot \bigg( \lambda_n \cdot \left(\frac{E}{\mathrm{GeV}}\right) \bigg)^n~\quad, \label{eq.Deltaphi} \\[0.35cm]
     &&  \mathrm{with:} \nonumber\\[0.35cm]
c_1 &=& \xi_1 \cdot \frac{d_\mathrm{Crab}}{c\cdot P_\mathrm{Crab}} \cdot 10^{-19}  \quad (\mathrm{GeV}^{-1}) \\[0.3cm]
c_2 &=& \xi_2 \cdot \frac{3}{2}\frac{d_\mathrm{Crab}}{c\cdot P_\mathrm{Crab}} \cdot 10^{-24}  \quad (\mathrm{GeV}^{-2})  \qquad,
\end{eqnarray}
where $d_\mathrm{Crab}$ is the pulsar distance, $c$ the Lorentz-invariant speed of light, $P_\mathrm{Crab}$ the pulsar period. 

We use now the profile likelihood ratio method~\cite{murphy2000} to define a test statistic $D_n$: 
\begin{equation}
D_n(\lambda_n|{\textit{\textbf X}}) = -2 \ln \left(\frac{\;\mathcal{L}(\lambda_n;\widehat{\widehat\bonu}(\lambda_n)|{\textit{\textbf X}})}
                                                  {  \mathcal{L}(\widehat{\lambda}_n; \widehat\bonu    |{\textit{\textbf X}})}\right) \quad. \label{eq.lr}
\end{equation}
of our pulsar dataset ${\textit{\textbf X}}=\{\epr_i,\phipr_i,k_i\}$ and a set of nuisance parameters $\bonu$. 
Here, $\epr_i$ is the reconstructed energy of each event $i$, $\phipr_i$ its reconstructed 
phase and $k_i$ the observation period. 
Single-hatted parameters $\{\widehat{\lambda}_n,\widehat\bonu\}$ maximize the likelihood, while 
double-hatted parameters $\widehat{\widehat\bonu}$ are those that maximize $\mathcal{L}$ under the assumption of $\lambda_n$.

The likelihood $\mathcal{L}$ takes the form of an \textit{extended likelihood}~\cite{barlow1995}:
\begin{eqnarray}
 \mathcal{L}(\lambda_n;\bonu|\textit{\textbf X}) &=& \mathcal{L}(\lambda_n;f,\alpha,\phip,\sigmap|\{\{\epr_i,\phipr_i\}_{i=0}^{N_k}\}_{k=0}^{N_s})  \\[0.3cm]
%{}  &=& \prod_{k=0}^{N_s}  \frac{(g_k(\lambda_n;\bonu)+b_k/\tau)^{N_k}}{N_k!} e^{-(g_k(\lambda_n;\bonu)+b_k/\tau)} 
%  \prod_{i=0}^{N_k} \; \mathcal{P}_k(\epr_i,\phipr_i|\lambda_n;\bonu) \cdot P(\bonu) ~,  \nonumber\\
%{}  &=& P(\bonu) \cdot \prod_{k=0}^{N_s} \frac{e^{-b_k}}{N_k^\textit{OFF}!}  \frac{e^{-(g_k(\lambda_n;\bonu)+b_k/\tau)}}{N_k^\textit{ON}!} 
{}  &=& P(\bonu) \cdot \prod_{k=0}^{N_s} \exp\left(-g_k(\lambda_n;\bonu)- b_k \cdot \frac{1 + \tau }{\tau}\right) \cdot \prod_{m=0}^{N_k^\textit{OFF}} b_k \cdot \nonumber\\
{}   &&  \cdot \prod_{i=0}^{N_k^\textit{ON}} \; \left(g_k(\lambda_n;\bonu)+b_k/\tau\right) \cdot \mathcal{P}_k(\epr_i,\phipr_i|\lambda_n;\bonu) ~. \label{eq.L} 
\end{eqnarray} \noindent
Here, $N_s$ denotes the number of observation periods, 
$N_k^\textit{ON}$ and $N_k^\textit{OFF}$ the number of events in the \ptwo \textit{ON} pulse region and the background control \textit{OFF} regions 
for observation period $k$, while $g_k$ and $b_k$ are their expectation values, respectively. We used $\phipr \in [0.3558,0.4495]$ to define the \textit{ON} region (optimized through simulations), 
$\phipr \in [0.52,0.87]$~\cite{fierro1998} for the \textit{OFF} region, and $\epr \in [0.4,7]$~TeV. 
This choice for the \textit{ON} region excludes contributions of \pone and practically all possible contributions from bridge emission~\cite{bridge14}. 
unnecessarily complicating the PDF and adding systematic uncertainties to the results. 
The ratio of phase width of the \textit{OFF}, divided by the one of the \textit{ON} region is labelled $\tau$. 
%Note the dependency of $g_k$ on both the LIV parameter $\lambda_n$ and all nuisance parameters, which are a direct consequence of the limited observed phase range. 
The background expectation values $b_k$ are direct nuisance parameters, while the signal expectation contains the flux normalization $f$ as nuisance parameter. 
A possible probability density function (PDF) for the nuisance parameters, known from external measurements, is labelled $P(\bonu)$.
The set of nuisance parameters contains, apart from the $b_k$: the \ptwo flux normalization $f$, its spectral index $\alpha$, 
the mean pulse position $\phip$ and its width $\sigmap$\footnote{% 
Nuisance parameters might also include additional asymmetry parameters, a spectral cutoff or other variables parameterizing a different pulse model.}.

The PDF of event $i$ is a combination of PDFs for signal (a pulsar event: $S_k(\epr_i,\phipr_i|\lambda_n;\bonu)$), 
or  the (interpolated) spectral energy distribution of the background: $h_k(\epr_i)$~(see e.g.~\cite{Segue2014}),
for the $k$-th data subsample, respectively:
\begin{eqnarray}
  \mathcal{P}_k(\epr_i,\phipr_i|\lambda_n;\bonu) &=& \frac{b_k/\tau \cdot h_k(\epr_i)~\, + ~\, g_k(\lambda_n;\bonu) \cdot S_k(\epr_i,\phipr_i|\lambda_n;\bonu)}{g_k(\lambda_n;\bonu)~\, + ~\, b_k/\tau} \quad. \label{eq.pdf}
\end{eqnarray}\noindent
The normalization constants of $S_k$ and $h_k$, and later $g_k$ and $b_k$, depend on all nuisance parameters and on $\lambda_n$. 
% once these PDFs are integrated within the phase window limits $\phipr_\textit{min}$ and $\phipr_\textit{max}$, and the reconstructed energy limits $\epr_\textit{min}$ and $\epr_\textit{max}$.
The signal PDF, $S_k(\epr_i,\phipr_i|\lambda_n;\bonu)$, is written as:  %$B_k$ is obtained from a control region $(0.52-0.87)$ in phase, considered as off-pulse region~\cite{fierro1998}.  $S_k$ is the corresponding spectral and phase distribution of the pulsar signal.: 
\begin{eqnarray}
S_k(\epr_i,\phipr_i|\lambda_n;\bonu) &=& \frac{\Delta t_k \int_0^\infty \!\!  R_k(E|\epr_i)\cdot \Gamma_{\ptwo}(E,f,\alpha) \cdot F_{\ptwo}(\phipr_i,E|\lambda_n;\phip,\sigmap) \; \diff E}{g_k(\lambda_n;\bonu)} ~.\noindent
\end{eqnarray}
Here, $\Delta t_k$ denotes the effective observation time for each $k$-th data subsample, 
$R_k$ the product of the effective collection area and the (inverted) energy re-distribution function to obtain a photon of true energy $E$, 
given its reconstructed energy $\epr$, both obtained from Monte-Carlo simulations.  
The \ptwo pulsar spectrum $\Gamma_{\ptwo}$ has been chosen to:
\begin{equation}
%\frac{\diff N}{\diff E\diff A \diff t}
\Gamma_{\ptwo}(E)
= f \cdot \big(E/E_\mathrm{dec}\big)^{-\alpha} \cdot \exp(-E/E_b)\quad\mathrm{TeV}^{-1}\,\mathrm{cm}^{-2}\,\mathrm{s}^{-1}\quad,  \label{eq.flux}
\end{equation}
%& a power-law with spectral index $-3.1$.
according to the findings of~\cite{crabtev}\footnote{%
\protect\cite{crabtev} excludes a possible spectral cutoff below 700~GeV.}. %, its location requires future data to be better constrained.
%($f_0 = (5.7\pm0.6)\cdot 10^{-10}~\mathrm{TeV}^{-1}\,\mathrm{cm}^{-2}\,\mathrm{s}^{-1}$, 
%a de-correlation energy~\cite{Fermidecorr2010} $E_\mathrm{dec} = 50\,\mathrm{GeV}$ and $\alpha = (3.0\pm 0.1)$ in a joint fit with \textit{Fermi} data, using a pure power-law (i.e. $E_b := \infty$).
The pulsar phaseogram model $F_\ptwo$ is computed as:
\begin{eqnarray}
F_\ptwo(\phipr_i,E|\lambda_n;\phip,\sigmap) \!
%                  &=& \!\int_0^\infty \!\!\!\! \frac{1}{2\pi\sigma_\textit{res}\sigma^\prime_\ptwo}
%                    \!\cdot\!
%                    \exp\!\bigg[\!- \!\frac{\Big(\phi_i\! -\! \phip\! -\! \Delta\phi(E|\lambda_n)\Big)^2}{2\,(\sigma^\prime_\ptwo)^2}\! -\! \frac{\Big(\phipr_i\! -\! \phi \Big)^2}{\,2\sigma^2_\textit{res}} \bigg]  \diff \phi \nonumber\\
                  &=& \frac{1}{\sqrt{2\pi}\sigmap}
                    \cdot
                    \exp\bigg[-\frac{\Big(\phipr_i - \phip - \Delta\phi(E|\lambda_n)\Big)^2}{2\,\sigmap^2} \bigg] \qquad, \label{eq:sympulse}
\end{eqnarray}
where the observed width $\sigmap$ contains contributions of the intrinsic pulse width and the instrumental phase resolution, both considered Gaussian 
in nature\footnote{%
An intrinsic Lorentzian pulse shape has been investigated as well, yielding similar results.},
%ehich is dominated by the uncertainties of the pulsar ephemerides, the RMS of the timing noise and the uncertainties of the barycentric corrections. 
%Since the latter contribution is two orders of magnitude smaller than the first~\cite{phd.garrido}, can be considered completely dominated by the intrinsic pulse width. 
%Note that the pulse form does not necessarily need to follow a Gaussian, and other, even asymmetric, functions cannot be excluded so far. The effect of different alternative possibilities will be investigated later on (see Section~\ref{sec.uncertainties}).
   %$F_{\mathrm{P2}} = \text{Gaus}(\phi_{\mathrm{P2}} + \Delta\phi_n,\sigma_{\mathrm{P2}})$
%	A Gaussian curve with a fixed width $\sigma_{\mathrm{P2}}$ was used,
and $\Delta\phi$ denotes the hypothetical phase delay produced by LIV, Eq.~\ref{eq.Deltaphi}.

\section{Results}

\begin{figure}
\centering
 \includegraphics[width=0.495\textwidth]{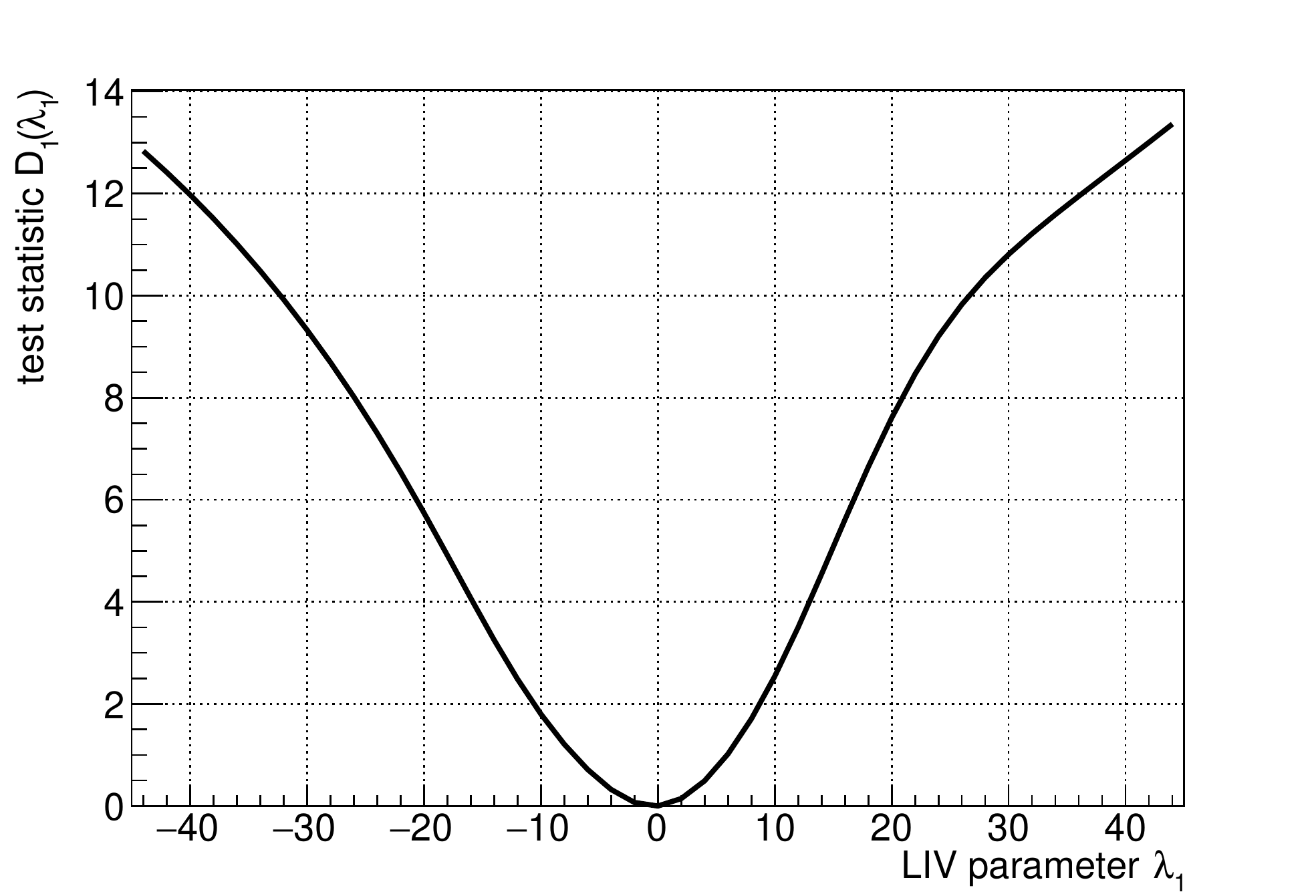}
 \includegraphics[width=0.495\textwidth]{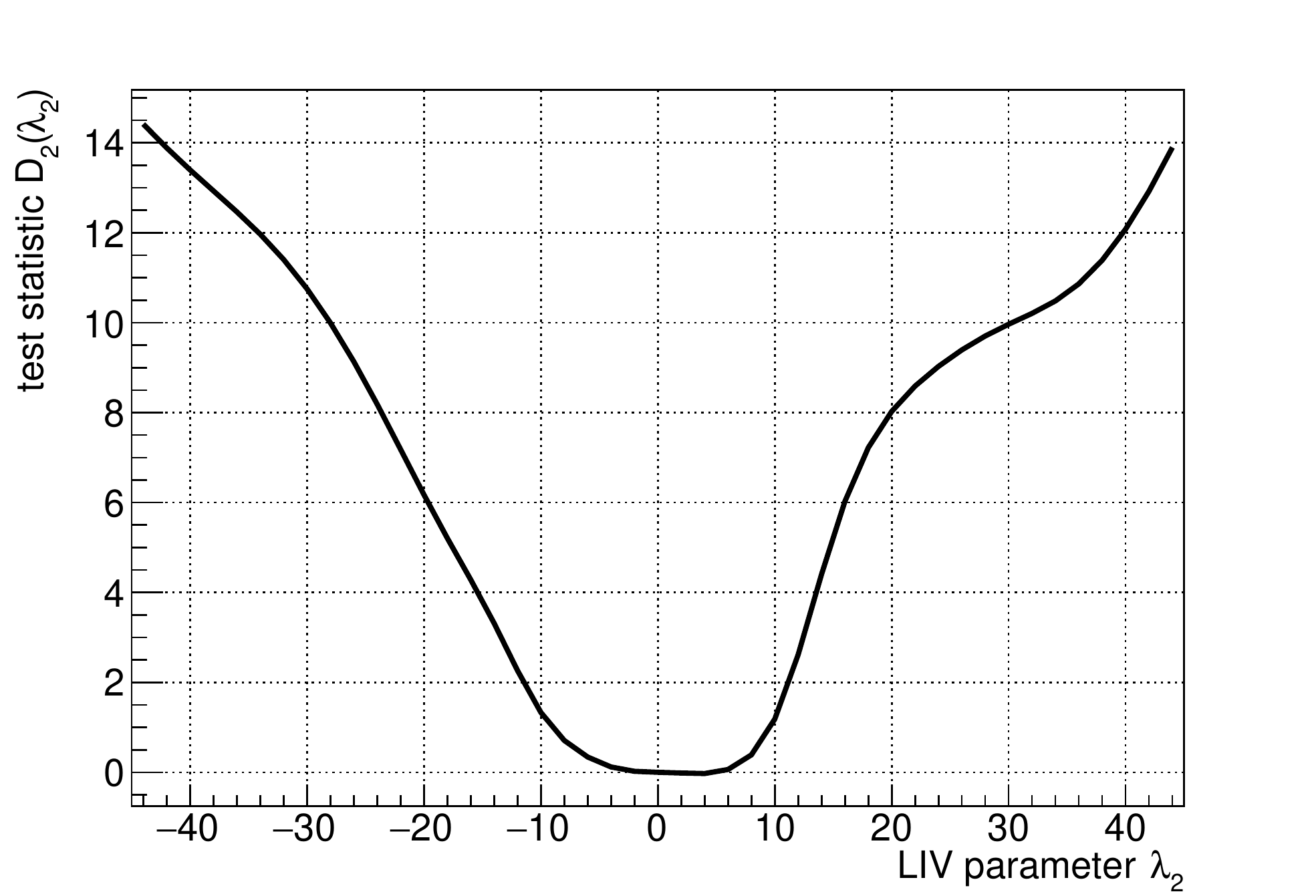}
  \caption{Left: test statistic (Eq.~\protect\ref{eq.lr}) as a function of  $\lambda_1$, right: as a function of $\lambda_2$.
    \label{fig7}}
\end{figure}

The profile likelihood algorithm Eq.~\ref{eq.lr} has been applied to the MAGIC Crab Pulsar data set~\cite{crabtev}, 
using the  \textit{TMinuit} class of \textit{ROOT}~\cite{tminuit} for the minimization.
The minima of the profile likelihood were found close to zero in all cases (see Fig.~\ref{fig7}).
Table~\ref{tab.MLnuisance} displays the nuisance parameters obtained at the 
minimum, all compatible with those obtained in~\cite{crabtev}. 
\begin{table}[h!]
\centering
\begin{tabular}{ccc}
\toprule
nuisance  &  result & unit   \\
 parameter &   \\%&  ($\epr_\textit{min} = 150$~GeV)  \\
\midrule
%%%            table2.C(X,1,1)             %%%            table2.C(X,1,1)
$\widehat{f}$  &  {\small $6.3   \pm 0.7 $}  & {\scriptsize $(\cdot 10^{-10}~\mathrm{TeV}^{-1}\,\mathrm{cm}^{-2}\,\mathrm{s}^{-1})$}  \\[0.01cm] % &   ${\small \boldsymbol{4.25   \pm 0.50  \pm 0.03 }}$   \\[0.01cm]
$\widehat{\alpha}$  & {\small $2.81  \pm 0.07   $} &  1 \\ %& ${\small \boldsymbol{2.41 \pm 0.06  \pm 0.03 }}$     \\%[0.2cm]
$\hphip$   & {\small $0.403 \pm 0.003  $} & 1 \\  %& {\small $0.4009 \pm 0.0028  \pm 0.0003 $}      \\%[0.2cm]
$\hsigmap$ & {\small $0.015 \pm 0.003  $} & 1 \\  %& ${\small \boldsymbol{ 0.026 \pm 0.004  \pm 0.001 }}$      \\
\bottomrule
\end{tabular}
\caption{Nuisance parameter values at the minima of $\lambda_{1,2}$. 
Uncertainties are statistical only, obtained from the 
diagonal elements of the covariance matrix, provided by \textit{TMinuit}. 
\label{tab.MLnuisance}}
\end{table}

\section{Discussion and Conclusions}

95\% confidence limits (CL) limits have been obtained by evaluating the likelihood at $\tilde{D}_n = \Delta\tilde{D}^{95\%}_n$, where 
$\Delta\tilde{D}^{95\%}_n$ has been obtained from simulations and found slightly higher than the canonical value of $\Delta\tilde{D}^{95\%}_n = 2.705$~\cite{pdg}. 
The difference is due to the nuisance parameters which have been varied in the simulations, using the covariance matrix obtained from the likelihood applied 
to experimental data. 

We studied systematic effects due to the insufficient knowledge, i.e. modelling of the likelihood, with respect to the background 
estimation, possible shifts in the assumed scale for energy and flux, different pulse shapes, including asymmetric behaviour, 
different values for the cutoff energy $E_b$, possible residual contributions from bridge emission and the uncertainty of the pulsar distance.
All values add up quadratically to about 42\% for the linear case, and 36\% for the quadratic one, respectively. 
It should be admitted here however, that energy dependent source-intrinsic effects might be possible and require further attention in the future.

Table~\ref{tab.MLlimits} shows the obtained limits, with and without systematic uncertainties. 

\begin{table}[h!]
\centering
%\begin{tabular}{ccc}
%\toprule
%case  &  limit (full range) & limit ($E\prime>400$~GeV)  \\
%\midrule
% %%% from figure8.C(1)
% $\xi_1 = +1$  & $E_{\textit{QG}_1} > X.X \cdot 10^{17}\text{ GeV }$ & $E_{\textit{QG}_1} > 8.9 \cdot 10^{17}\text{ GeV }$  \\
% $\xi_1 = -1$  & $E_{\textit{QG}_1} > X.X \cdot 10^{18}\text{ GeV }$ & $E_{\textit{QG}_1} > 6.8 \cdot 10^{17}\text{ GeV }$ \\
% %%% from figure8.C(2)
% $\xi_2 = +1$  & $E_{\textit{QG}_2} > X.X \cdot 10^{10}\text{ GeV }$ & $E_{\textit{QG}_2} > 6.2 \cdot 10^{10}\text{ GeV }$ \\
% $\xi_2 = -1$  & $E_{\textit{QG}_2} > X.X \cdot 10^{10}\text{ GeV }$ & $E_{\textit{QG}_2} > 4.5 \cdot 10^{10}\text{ GeV }$ \\
\begin{tabular}{ccc}
  \toprule
case  &  95\% CL limit (w/o systematic) & 95\% CL limit (incl. systematics) \\
\midrule
%%% from figure8.C(5)
$\xi_1 = +1$  & $E_{\textit{QG}_1} > 7.8 \cdot 10^{17}\text{ GeV }$ & $E_{\textit{QG}_1} > 5.5 \cdot 10^{17}\text{ GeV }$  \\
$\xi_1 = -1$  & $E_{\textit{QG}_1} > 6.4 \cdot 10^{17}\text{ GeV }$ & $E_{\textit{QG}_1} > 4.5 \cdot 10^{17}\text{ GeV }$  \\
%%% Figure7/ScanLIV_TeVPaper_Sigx1EnCut400-7TeV_Cutoff1000.0TeV_0_0_0_0_0_1_1_1_1_1_0_0_0_excl20_Ph0.3558-0.4495_WithPrior_Profiled4_TestLivOrder2.root
%%% TF1 *f2 = new TF1("f","[1]*(x-[0])*(x-[0])+[2]",-20.,25.)
%%% hLivScan->Fit(f2,"R","same",-20.,18.)
%%% // f2->Eval(-16.7)
%%% f2->Eval(-13.94)
%%% (Double_t) 2.76169
%%% hLivScan->Fit(f2,"R","same",0.,18.)
%%% // f2->Eval(13.73)
%%% %%% f2->Eval(12.47)
%%% (Double_t) 2.76117
%%% .L figure8.C
%%% // LivParam2LogEQG(13.73,2)
%%% LivParam2LogEQG(12.47,2)
%%% (Double_t) 8.01925e+10
%%% // LivParam2LogEQG(16.7,2)
%%% LivParam2LogEQG(13.94,2)
%%% (Double_t) 7.17360e+10
%%% 
%%%
%%% Likelihood: 
%%% Quadratic LIV factor = 0.00 ; Mean P2 pos = 0.402703 ; Width: 0.018272, F_0: 6.81487, alpha: 2.75565, WidthE: 0.00000, Ecut: 999999.00, Width2: 0.014
%%% 819, PhiDiff: 0.00000000, Ephidiff: 999999.000000, -2 loglike = 2( -75544.778877503806143 - -75544.479362969053909 ) = -5.99029e-01 
%%% -2logL = -0.599029; iflag = 0 (converged)
%$\xi_2 = +1$  & $E_{\textit{QG}_2} > 7.6 \cdot 10^{10}\text{ GeV }$ &  $E_{\textit{QG}_2} > 5.9 \cdot 10^{10}\text{ GeV }$  \\
%$\xi_2 = -1$  & $E_{\textit{QG}_2} > 5.9 \cdot 10^{10}\text{ GeV }$ &  $E_{\textit{QG}_2} > 4.5 \cdot 10^{10}\text{ GeV }$ \\
$\xi_2 = +1$  & $E_{\textit{QG}_2} > 8.0 \cdot 10^{10}\text{ GeV }$ &  $E_{\textit{QG}_2} > 5.9 \cdot 10^{10}\text{ GeV }$  \\
$\xi_2 = -1$  & $E_{\textit{QG}_2} > 7.2 \cdot 10^{10}\text{ GeV }$ &  $E_{\textit{QG}_2} > 5.3 \cdot 10^{10}\text{ GeV }$ \\
\bottomrule
\end{tabular}
\caption{Obtained limits applying the profile likelihood method. \label{tab.MLlimits}}
\end{table}

These limits are found well below experimental results obtained on GRBs~\cite{fermiliv13} and hence not competitive for the linear case. 
The quadratic case yields, however, constraints only about a factor two from the current best limits~\cite{fermiliv13}
% is $E_{\textit{QG}_2} > 8.6 \cdot 10^{10}$~GeV ($E_{\textit{QG}_2} > 13 \cdot 10^{10}$~GeV if 
%the ``Sharpness-maximization method'' is used instead of likelihood) for the subluminal case, and $E_{\textit{QG}_2} > 9.4 \cdot 10^{10}$~GeV for the superluminal case~\citep{fermiliv13}, 
%are only about a factor of two better than our limits. %, and an improved limits are 
%The advantage of using Crab pulsar data to limit LIV is nevertheless that 
%the limit can be systematically improved over time, as more and more data are taken. 
Since there are currently strong arguments against linear LIV effects, even suppressed by the Planck energy~\cite{Kislat:2017,Goetz:2014}, 
limits constraining the quadratic case are now of greater interest. 
%Moreover, if LIV is not isotropic, 25 non-birefringent coefficients must be constrained via direction dependent limits~\cite{Kislat:2015}, 
%whose current best values are five to six orders of magnitude worse.

%Unlike flaring astrophysical sources like AGNs or GRBs, 
Pulsar data has the advantage that it can be continuously accumulated and sensitivity to LIV improved.
%The likelihood is currently still dominated by background fluctuations as well as un-resolved systematics, particularly the 
%pulse shape and its evolution with energy. Such effects have possibly been found between 200 and 300~GeV, although the given statistics does not allow to claim 
%firm detection. More data will eventually allow to shed light on the pulse evolution in this energy range and 
%subsequently include events with reconstructed energies below 400~GeV into the likelihood analysis. 
%Our simulations (see Fig.~\ref{fig.figure4b}) have shown that this possibility alone \refonecol{may} already improve the limits by at least a factor of two. 
%Moreover, more data will allow to better model the pulse shape itself and take less conservative choices than the used  
% Gaussian pulse shape with fixed symmetric width. 
MAGIC is currently at the zenith of its performance~\cite{perform16}, which gives the possibility to take regular data on the Crab Pulsar, 
particularly at higher zenith angles where sensitivity for TeV energy gamma-rays is better. We expect that a 
data set of 2000~hours of stereo data, a number within reach for the MAGIC collaboration, 
%given the regular observation of the Crab Pulsar for calibration purposes, 
can ensure an improvement of the quadratic limit by a factor of two, most probably even exceeding the current Fermi limit~\cite{fermiliv13}.
Moreover, our profile likelihood can be combined with that from other sources, like AGNs, or even other experiments. 
In such a way, significantly improved constraints on LIV are well within reach in the next years.

% Within a framework of collaboration between the different current IACT installations, a significantly higher amount of data is even plausible.
%Such a limit will reach the current world-best constraints, but has the possibility to go well beyond these, since these data can also help to better understand
%pulse evolution of the inter-pulse and such allow to include events below 400~GeV in the likelihood.

%Moreover, it is hoped that the Gaia mission~\cite{GaiaMission:2016}
%will soon be able to measure the distance to Crab to at least an order of magnitude better precision, removing one of the main uncertainties to these limits.

\acknowledgments

We would like to thank the IAC for the excellent working conditions at the ORM in La Palma. We acknowledge the financial support of the German BMBF, DFG and MPG, 
the Italian INFN and INAF, the Swiss National Fund SNF, the European ERDF, the Spanish MINECO, the Japanese JSPS and MEXT, the Croatian CSF, and the Polish MNiSzW.

\end{document}